**Joint longitudinal and time-to-event models for multilevel hierarchical data**

Samuel L Brilleman[1,2] *

Michael J Crowther[3]

Margarita Moreno-Betancur[2,4,5]

Jacqueline Buros Novik[6]

James Dunyak[7]

Nidal Al-Huniti[7]

Robert Fox[7]

Jeff Hammerbacher[6,8]

Rory Wolfe[1,2]

**Author Affiliations:** [1] Department of Epidemiology and Preventive Medicine, School of Public Health and Preventive Medicine, Monash University, Melbourne, Australia; [2] Victorian Centre for Biostatistics (ViCBiostat), Melbourne, Australia; [3] Biostatistics Research Group, Department of Health Sciences, University of Leicester, Leicester, UK; [4] Clinical Epidemiology and Biostatistics Unit, Murdoch Children's Research Institute, Melbourne, Australia; [5] Melbourne School of Population and Global Health, University of Melbourne, Melbourne, Australia; [6] Department of Genetics and Genomic Sciences, Icahn School of Medicine at Mount Sinai, New York, NY, USA; [7] Quantitative Clinical Pharmacology, IMED Biotech Unit, AstraZeneca, Waltham, MA, USA; [8] Department of Microbiology and Immunology, Medical University of South Carolina, Charleston, SC, USA;

**\* Corresponding author:**

Postal: School of Public Health and Preventive Medicine, Monash University, 553 St Kilda Road, Melbourne, VIC 3004, Australia

Phone: +61 (4) 9903 0802

Email: sam.brilleman@monash.edu

**Acknowledgements**: SLB is funded by an Australian National Health and Medical Research Council (NHMRC) Postgraduate Scholarship (ref: APP1093145), with additional support from an NHMRC Centre of Research Excellence grant (ref: 1035261) awarded to the Victorian Centre for Biostatistics (ViCBiostat). MJC is partly funded by a UK Medical Research Council (MRC) New Investigator Research Grant (ref: MR/P015433/1). We thank Eric Novik at Generable (http://www.generable.com/) for his support and management of the collaboration between SLB, JBN and the team at AstraZeneca (NAH, RH, JD).

**Conflicts of interest**: None.

**Supplementary Materials**: We have developed user-friendly software for fitting the model described in the paper. The software is publically available as part of the rstanarm package, downloadable from the Comprehensive R Archive Network (https://cran.r-project.org/). The supplementary materials include an example of the code required to fit the model and additional details about the model estimation. However, the Iressa Pan-Asia Study (IPASS) dataset used in our application is not publicly available.

**Abstract**

Joint modelling of longitudinal and time-to-event data has received much attention recently. Increasingly, extensions to standard joint modelling approaches are being proposed to handle complex data structures commonly encountered in applied research. In this paper we propose a joint model for hierarchical longitudinal and time-to-event data. Our motivating application explores the association between tumor burden and progression-free survival in non-small cell lung cancer patients. We define tumor burden as a function of the sizes of target lesions clustered within a patient. Since a patient may have more than one lesion, and each lesion is tracked over time, the data have a three-level hierarchical structure: repeated measurements taken at time points (level 1) clustered within lesions (level 2) within patients (level 3). We jointly model the lesion-specific longitudinal trajectories and patient-specific risk of death or disease progression by specifying novel association structures that combine information across lower level clusters (e.g. lesions) into patient-level summaries (e.g. tumor burden). We provide user-friendly software for fitting the model under a Bayesian framework. Lastly, we discuss alternative situations in which additional clustering factor(s) occur at a level *higher* in the hierarchy than the patient-level, since this has implications for the model formulation.

1. Introduction

In clinical or epidemiological research studies, longitudinal data may be in the form of a clinical biomarker that is repeatedly measured over time on a given patient, whilst time-to-event data may refer to the patient-specific time from a defined origin (e.g. time of diagnosis of a disease) until a clinical event of interest such as death or disease progression. A common motivation for collecting such data is to explore how changes in the biomarker are associated with the occurrence of the event. A rapidly evolving field of statistical methodology, known as "joint modelling", aims to model both the longitudinal and time-to-event data simultaneously providing several potential benefits over more traditional approaches [1–3]. Compared with using the observed biomarker measurements as covariates in a time-to-event model, a joint modelling approach can protect against bias due to missing data or measurement error in estimating the association between the value of the biomarker and the risk of occurrence of the event [1,4]. Moreover, we can explore the associations between more complex aspects of the biomarker trajectory (such as the rate of change) and the occurrence of the event. Lastly, we might wish to use the longitudinal biomarker data in the development of a "dynamic" risk prediction model, and joint modelling approaches lend themselves to this purpose [5,6].

The so-called "shared parameter" joint modelling approach consists of two regression submodels, one for the longitudinal biomarker measurements (the "longitudinal submodel") and one for the time-to-event outcome (the "event submodel"). Dependence between the two submodels is allowed for by assuming that the model for the time-to-event outcome includes as predictor some functional form of the patient-specific parameters from the longitudinal submodel, commonly referred to as an association structure. In the joint modelling literature to date, primary focus has been on a situation in which there is a single normally-distributed biomarker measured repeatedly over time for each patient and a unique, possibly right-censored, time to a terminating event of interest [4,7]. However, a number of extensions have been proposed for the standard shared parameter joint model, such as competing risks [8], interval censored event times [9], non-normally distributed biomarkers [10], and multiple biomarkers [11].

Nonetheless, a common aspect of the proposed methodology has been that the longitudinal data have a two-level hierarchical structure; longitudinal measurements of the biomarker are observed at time points (level 1 of the hierarchy) which are clustered within patients (level 2 of the hierarchy). The patient is therefore considered to be the only clustering factor. An example of this data structure is shown in Figure 1a. However, there exist many situations in clinical and epidemiological research in which we wish to analyse longitudinal and time-to-event data where the longitudinal data component (and potentially also the time-to-event component) has a hierarchical structure with clustering factors beyond just that of the patient.

In this paper we describe a joint modelling approach that can be applied to longitudinal and time-to-event data with more than one clustering factor. In Section 2 we introduce several motivating examples which describe the types of data structures our joint modelling approach is intended for. In Sections 3 and 4 we describe the formulation and estimation of a joint model that is suitable when an additional clustering factor occurs at a level *lower* in the hierarchy than the patient-level. In Section 5 we describe an application in which we use this joint model to explore the association between tumor burden and risk of death or disease progression in non-small cell lung cancer (NSCLC) patients undergoing treatment. In

Section 6 we describe the formulation of the joint model under alternative scenarios in which the additional clustering factor occurs at a level *higher* in the hierarchy than the patient-level. In Section 7 we close with a discussion.

2. Motivating examples

2.1 Tumor burden and progression-free survival in non-small cell lung cancer

In our primary motivating example interest lies in exploring the relationship between tumor burden and the risk of death or disease progression in patients with non-small cell lung cancer (NSCLC). After a patient initiates treatment the size of each tumor lesion is measured repeatedly over time in order to assess the effectiveness of treatment and aid clinical decision making. Accordingly, for a given patient, we can define the tumor burden as some patient-level summary of the sizes of their individual tumor lesions. Given that a patient may have more than one lesion, our data consists of a hierarchy in which the longitudinal measurements are observed at time points (level 1) which are clustered within a specific lesion (level 2) for a given patient (level 3), as represented in Figure 1b.

Consideration of the multilevel structure of the data is important for several reasons. Firstly, the underlying growth trajectories may vary across different lesions, even when those lesions are clustered within the same patient. We can allow for between-lesion variation in the growth trajectories through the use of lesion-specific, as well as patient-specific, parameters in the longitudinal submodel. Equivalently, the introduction of lesion-specific parameters in the longitudinal submodel allows us to account for the within-cluster correlation of longitudinal measurements made on the same lesion and therefore appropriately estimate standard errors. Secondly, the hierarchical structure of the data is a key aspect to consider when specifying the form of the association between the longitudinal and event processes, something we discuss further in Section 3.3.

2.2 Visual field progression in glaucoma

Our second motivating example comes from research on eye disease. In ophthalmology it is of interest to use repeated measurements of eye-specific biomarkers to help predict the occurrence of disease-specific events. For example, in glaucoma research we may be interested in the association between optic nerve head surface depth (ONHSD) and visual field progression. Previous studies [12] have used joint models to explore this association by treating each eye as independent and modelling the association between the eye-specific longitudinal trajectory for ONHSD and the eye-specific event endpoint (visual field progression). However, this approach ignores the dependence between measurements taken on the two eyes clustered within a patient. Arguably, a more appropriate analysis approach would model the correlation between measurements taken on a person's two eyes. Hence, consider a joint modelling approach in which we assume the ONHSD measurements are observed at time points (level 1) which are clustered within a specific eye (level 2) for a given patient (level 3). We could then explore the association between the longitudinal trajectory for ONHSD and a patient-specific endpoint for the time to visual field progression.

2.3 Patients within clinics or the meta-analysis of joint model data

Our final two motivating examples relate to an alternative data structure in which the additional clustering factor occurs at a level which is *higher* in the hierarchy than the patient. One example is where repeated observation times (level 1) exist for patients (level 2) and those patients are clustered within clinics (level 3). Another example is an individual patient data (IPD) meta-analysis where observation times (level 1) are for patients (level 2) clustered within randomised clinical trials (level 3) [13]. In both of these examples, we wish to include the additional clustering factor (i.e. the clinic or the trial) in our joint modelling approach, so that we appropriately allow for the correlation structure. However, because the additional clustering factor occurs at a level higher than the patient-level, there are different implications for the specification of the joint model association structure compared with our previous motivating examples. For this reason we describe a formulation of the joint model for this type of data structure separately; in Section 6 of the paper.

3. Model formulation

3.1 Longitudinal submodel

Consider the situation in which we have a three-level hierarchical structure for our longitudinal data, where the patient represents the highest level of the hierarchy (in Section 6 we discuss the situation in which the patient does not represent the highest level of the hierarchy). We assume our longitudinal outcome measurements $y_{ijk} = y_{ij}(t_{ijk})$ are obtained at a set of time points $k = 1, \ldots, K_{ij}$ which are assumed to be nested within unit $j$ ($j = 1, \ldots, J_i$) of the level 2 clustering factor which in turn is nested within patient $i$ ($i = 1, \ldots, N$), the level 3 clustering factor. We model the longitudinal outcome in continuous time using a generalised linear mixed effects model where we assume $Y_{ij}(t)$ is governed by a distribution in the exponential family with expected value $\mu_{ij}(t) = g^{-1}(\eta_{ij}(t))$ for some known link function $g(.)$. Specific choices of family and link function lead to, for example, linear, logistic or Poisson regression. We specify a three-level hierarchical model for the linear predictor

$$\eta_{ij}(t) = x'_{ij}(t)\beta + z'_{ij}(t)b_i + w'_{ij}(t)u_{ij} \qquad (1)$$

where $x_{ij}(t)$, $z_{ij}(t)$, and $w_{ij}(t)$ are vectors of covariates, possibly time-dependent. The vector $\beta$ contains fixed-effect parameters, and $u_{ij}$ and $b_i$ are vectors of level 2 (cluster-specific) and level 3 (patient-specific) parameters, each assumed to be normally distributed with mean zero and unstructured variance-covariance matrix, that is $u_{ij} \sim N(0, \Sigma_u)$ and $b_i \sim N(0, \Sigma_b)$. We assume that $u_{ij}$ and $b_i$ are uncorrelated.

3.2 Event submodel

We observe an event time $T_i = min\{T_i^*, C_i\}$, where $T_i^*$ denotes the true event time for patient $i$ and $C_i$ denotes the right-censoring time, and define an indicator of observed event occurrence $d_i = I(T_i^* \leq C_i)$. We model the hazard of the event using a proportional hazards regression model

$$h_i(t) = h_0(t) exp\left(v'_i(t)\gamma + \sum_{q=1}^{Q} \alpha_q f_q(\Theta_{ij}(t); j = 1, \ldots, J_i)\right) \qquad (2)$$

where $h_i(t)$ is the hazard of the event for patient $i$ at time $t$, $h_0(t)$ is the baseline hazard at time $t$, $v_i(t)$ is a vector of covariates with an associated vector of fixed-effect parameters $\gamma$, and $\sum_{q=1}^{Q} \alpha_q f_q(\Theta_{ij}(t); j = 1, \ldots, J_i)$ forms the "association structure" for the joint model which consists of some specified set of functions $f_q(.)$ applied to the full set of (possibly time-varying) parameters from the longitudinal submodel $\Theta_{ij}(t) = \{\beta, b_i, u_{ij}, \mu_{ij}(t), \eta_{ij}(t)\}$ with associated fixed effects $\alpha_q$ ($q = 1, \ldots, Q$). The functions $f_q(.)$ might each correspond to a functional of the longitudinal submodel parameters for a given patient $i$ and cluster $j$, for example, the expected value or rate of change in the longitudinal biomarker. Alternatively, they might be functions of the longitudinal submodel parameters for a given patient $i$ across all $J_i$ clusters, representing different methods for combining the level 2 clusters into a patient-level summary (as described in the next section). We refer to the fixed effects $\alpha_q$ as

"association parameters" since they quantify the magnitude of the association between aspects of the longitudinal process and the event process. In the next section we describe the variety of ways in which the association structure for the joint model can be specified.

3.3 Association structures for patient-level summaries

Given that the event time $T_i$ is measured at the patient-level, the patient represents the level of the hierarchy at which our primary interest lies for understanding the association between the longitudinal and event processes. Accordingly, we wish to formulate a model that captures the association between the longitudinal and event processes at any given time $t$ in a meaningful way at the patient-level. A decision is required about how information from the level 2 clustering factor (that is, the clustering factor between the patient-level and the observation-level) is used in the formulation of the association structure.

Since the number of level 2 units may differ for each patient (i.e. it isn't necessarily the case that $J_i = J_{i'}$ for all $i \neq i'$) we must combine the information in the level 2 units into some patient-level time-specific summary. Obvious choices for a patient-level summary measure are likely to be the summation, average, maximum or minimum taken across the level 2 units within patient $i$. That is

$$f_q(\Theta_{ij}(t); j = 1, \dots, J_i) = \sum_{j=1}^{J_i} \mu_{ij}(t) \tag{3}$$

$$f_q(\Theta_{ij}(t); j = 1, \dots, J_i) = J_i^{-1} \sum_{j=1}^{J_i} \mu_{ij}(t) \tag{4}$$

$$f_q(\Theta_{ij}(t); j = 1, \dots, J_i) = max\ (\mu_{ij}(t);\ j = 1, \dots, J_i) \tag{5}$$

$$f_q(\Theta_{ij}(t); j = 1, \ldots, J_i) = min\,(\mu_{ij}(t);\ j = 1, \ldots, J_i) \quad (6)$$

The association structure resulting from equation (3) assumes that the hazard of the event for patient $i$ at time $t$ is associated with the sum of the expected values (at time $t$) for each of the level 2 units clustered within that patient. In contrast, the $J_i^{-1}$ term in equation (4) provides us with the average of the level 2 cluster-specific expected values within patient $i$ rather than their summation alone. Lastly, equations (5) and (6), respectively, assume that the hazard of the event for patient $i$ at time $t$ is associated with the level 2 cluster (within patient $i$) that has the largest or smallest expected value at time $t$. It is possible for more than one of these summary functions to be included in a single model (i.e. $Q > 1$). Moreover, other summary functions are possible but are not described here.

The most appropriate summary function(s) may be determined based on clinical context, or by choosing the association structure that provides the best model performance based on some criterion. For instance, returning to the first motivating example introduced in Section 2, we may believe that risk of death or disease progression for a patient with NSCLC is driven by treatment-failure occurring at a single lesion. We may therefore assume the hazard of the event for patient $i$ at time $t$ is associated with the maximum (i.e. largest) of the lesion-specific expected values, since this would represent the lesion with the most advanced disease, for example due to it having the worst treatment response.

Moreover, we could easily replace the $\mu_{ij}(t)$ in equations (3) through (6) with some other function of the longitudinal submodel parameters, such as the level 2 cluster-specific rate of change in the marker at time $t$ (i.e. $\frac{d\mu_{ij}(t)}{dt}$) or the area under the level 2 cluster-specific marker trajectory up to time $t$ (i.e. $\int_0^t \mu_{ij}(u)\,du$). For instance, we may assume that the lesion with largest growth rate may be most informative of treatment failure. Such extensions follow naturally from association structures that have been proposed elsewhere for shared parameter joint models [11,14].

The specifications in equations (3) and (4) both assume a constant magnitude of association between the expected value of each level 2 unit and the hazard of the event; that is, there is an implicit assumption that the level 2 units within a patient are exchangeable since their expectations are each multiplied by the same fixed effect association parameter $\alpha_q$. It is worth noting that this is in contrast to the situation in which we have multiple longitudinal biomarkers (for example lesion size and circulating DNA) each measured repeatedly over time. In this situation the multiple biomarkers within a patient are *not* exchangeable and therefore each biomarker would have a different coefficient quantifying its association with the hazard of the event. Methods for the joint modelling of multiple longitudinal biomarkers and time-to-event data, where the multiple biomarkers are not exchangeable, have been described elsewhere [11,15,16]. Although it is outside the scope of this paper, the methodology described here could be extended to a situation in which we have multiple longitudinal biomarkers, some of which may or may not have additional levels of clustering. This type of data structure is therefore represented in Figure 1c.

4. Model estimation

The joint model proposed in Section 3 can be estimated using the 'stan_jm' modelling function within the rstanarm R package [17]. The association structure can be based on the expected value and/or slope of the longitudinal biomarker, however, currently only a single patient-level summary function (summation, average, maximum, or minimum) can be chosen by the user (this restriction may be relaxed in a future release). Estimation of the model requires a full Bayesian specification with prior distributions on all unknown parameters. We provide further details of the estimation (for example prior distributions and computation) in the Supplementary Materials.

5. Application

We demonstrate the use of our modelling approach by exploring the association between tumor burden and the hazard of death or disease progression amongst NSCLC patients undergoing treatment.

5.1 Data

The Iressa Pan-Asia Study (IPASS) was an open label, phase 3 trial of 1,217 untreated NSCLC patients in East Asia randomized to: (i) gefitinib (250mg per day), or (ii) carboplatin (dose calculated to provide 5-6 mg per milliliter per minute) plus paclitaxel (200 mg per square meter of body-surface area) [18]. The primary endpoint was progression-free survival, however, the study was extended to track overall survival in the longer term. We restricted our analyses to the 430 (35%) patients with an available test result for epidermal growth factor receptor (EGFR) mutation since this has been shown to be associated with both tumor dynamics and treatment response [19]. We thereby defined a group covariate corresponding to either: (i) EGFR+, (ii) EGFR- and receiving gefitinib; or (iii) EGFR- and receiving carboplatin plus paclitaxel.

5.2 Model specification

5.2.1 Longitudinal submodel

We modelled repeated measurements of the longest diameter (in millimetres) of each lesion using a linear mixed effects model (identity link, normal distribution) with a linear predictor as in equation (1) where: the observation-level covariates with fixed (population-average) effects, $x_{ij}(t)$, included an intercept, the 3-category EGFR-group covariate, linear and quadratic terms for time, and an interaction between group and each of the linear and quadratic terms for time; the patient-level vector $z_{ij}(t)$ included an intercept only; and the lesion-level vector $w_{ij}(t)$ included an intercept, and linear and quadratic terms for time. This specification allowed for lesion-specific nonlinear (quadratic) evolutions of the longitudinal trajectory, while also allowing the average (i.e. population-level) estimate of the nonlinear longitudinal trajectory to differ between the three groups (through the group by time interaction terms).

### 5.2.2 Event submodel

We modelled the hazard of death or disease progression using the proportional hazards model in equation (2). We approximated the log baseline hazard using B-splines with 6 degrees of freedom and included a 3 category physical functioning measure (normal activity; restricted activity; in bed >50% of the time) [20] as a baseline covariate in $v_i(t)$. We considered several models which each differed in terms of their association structure. Specifically we considered the following: (i) no association structure (i.e. no biomarker information in the event submodel), (ii) association structures based on the sum, average, maximum or minimum of the lesion-specific expected values (i.e. the association structures defined in equations (3) through (6)), and (iii) association structures based on both the lesion-specific expected value *and* slope, that is an association structure of the form

$$\sum_{q=1}^{Q} \alpha_q f_q(\Theta_{ij}(t)) = \alpha_1 f(\mu_{ij}(t); j = 1, \ldots, J_i) + \alpha_2 f(\frac{d\mu_{ij}(t)}{dt}; j = 1, \ldots, J_i) \quad (7)$$

where the function $f(.)$ was taken to be either the sum, average, maximum or minimum, $\mu_{ij}(t)$ is the size and $\frac{d\mu_{ij}(t)}{dt}$ is the rate of change in the size of lesion $j$ in patient $i$ at time $t$, $J_i$ is the total number of target lesions identified for patient $i$ at baseline, and $\alpha_1$ and $\alpha_2$ are association parameters.

5.3 Model comparison

An ideal feature of our model would be that it is able to inform clinical decision making by accurately predicting a patient's future risk of death or disease progression in the clinical setting. We therefore compared different possible association structures for our proposed joint model using a measure of predictive accuracy for the event outcome. Specifically, we used the estimated area under the (time-dependent) receiver operating characteristic curve (AUC) to assess how well each of the models discriminated between those patients who did and did not have the event [3].

To do this we first used the fitted joint model to generate conditional survival probabilities for each patient at some time horizon $t_L + \Delta t$, conditional on: (i) their still being at risk at some landmark time $t_L$, and (ii) their longitudinal biomarker data up to the landmark time $t_L$ (following the methods described in [3]). These survival probabilities were then used in combination with the observed event times and censoring indicators for each patient, taken over the interval $(t_L, t_L + \Delta t)$, to calculate the time-dependent AUC measure.

5.4 Results

In our analysis, 360 (84%) of the 430 patients progressed or died prior to censoring. The overall Kaplan-Meier curve is shown in Figure 2. There were 1209 lesions across the 430 patients, and 138 (32%), 101 (23%), 71 (17%) and 120 (28%) patients with 1, 2, 3 or 4+ lesions, respectively. A total of 6132 size measurements were observed, corresponding to a median number of 5 (IQR: 3 to 7; range: 1 to 17) measurements per lesion.

Table 1 shows the estimated AUC values for the fitted models. The results are shown for a landmark time of $t$ = 5 months and a horizon time of $t + \Delta t$ = 10 months. For an association structure based on the expected value (i.e. diameter of the lesion) only, a summary function based on the sum or maximum of the lesions showed better discriminatory performance compared with using the mean or minimum. We found that also including the slope (i.e. rate of change in the diameter of the lesion) in the association structure improved the predictive performance. When both the expected value and slope were used in the association structure then summaries based on the sum, mean or maximum of the lesions all performed similarly. The summary based on the minimum (i.e. size of the smallest lesion, and rate of change in the slowest growing lesion, at time $t$) was the worst in terms of predicting the risk of death or disease progression. These results are in line with what we would expect from a clinical perspective, that is, those summaries that incorporate information on the largest and/or fastest growing lesion at time $t$ are likely to provide better predictive performance. This is because they capture information about the most aggressive tumor, which may have escaped treatment and is therefore likely to impact most severely on the risk of death disease progression and death.

Table 2 shows the parameter estimates from the model with the best performance based on the AUC measure (with an association structure based on the maximum of the lesion-specific expected values and slopes). The estimated hazard ratio corresponding to the first association parameter (i.e. $exp(\alpha_1)$) was 1.011 (95% credible interval (CrI): 1.004 to 1.017), suggesting that a one millimetre increase in the diameter of a patient's largest lesion was associated with a 1.1% (95% CrI: 0.4 to 1.7%) increase in their hazard of death or disease progression (conditional on the other covariates in the model). Similarly, a one millimetre per month increase in the rate of change of their fastest growing lesion was associated with a 56% (95% CrI: 42 to 75%) increase in their hazard. Figure 3 shows the fitted lesion-specific longitudinal trajectories and observed measurements for a selection of patients under the fitted model.

6. Alternative data structure: clustering *above* the patient-level

In our analysis of the IPASS data, patient represented the top level of the data hierarchy and the additional clustering factor – "lesion" – occurred at a level which was *lower* in the hierarchy than patient; that is, lesions were clustered within patients rather than patients being clustered within lesions. An alternative situation is that in which the additional clustering factor(s) occur at a level which is *higher* in the hierarchy than the patient-level. An example is where repeated observation times (level 1) exist for patients (level 2) and the patients are clustered within clinics (level 3). Another example is an individual patient data (IPD) meta-analysis with repeated observation times (level 1) for patients (level 2) clustered within randomised clinical trials (level 3) [13].

Recall however that the event time $T_i$ is measured at the patient-level and, therefore, the patient represents the level of the hierarchy at which our primary interest lies for understanding the association between the longitudinal and event processes. For this reason, the relative locations within the hierarchy of the patient and the additional clustering factor have implications for specifying the association structure of the joint model.

6.1 Model formulations based on a *patient*-level association structure

In Section 3.3 we proposed association structures based on a patient-level time-specific summary of the $J_i$ level 2 units clustered within patient $i$. However, with clustering above the patient-level, there is no need to construct such a patient-level summary.

Suppose that the longitudinal outcome $y_{lik} = y_{li}(t_{lik})$ is measured at time point $k$ ($k = 1, \ldots, K_{li}$) which is nested within unit $i$ ($i = 1, \ldots, N_l$) of the level 2 clustering factor (the patient) which in turn is nested within unit $l$ ($l = 1, \ldots, L$) of a level 3 clustering factor (clinic, say, for example). If we again model the longitudinal outcome in continuous time using a generalised linear mixed effects model where $Y_{li}(t)$ is governed by a distribution in the exponential family with expected value $\mu_{li}(t) = g^{-1}(\eta_{li}(t))$ we might, for example, consider a specification for the longitudinal submodel of the form

$$\eta_{li}(t) = x'_{li}(t)\beta + z'_{li}(t)b_{li} + q'_{li}(t)c_l \qquad (8)$$

where $x_{li}(t)$, $z_{li}(t)$ and $q_{li}(t)$ are vectors of covariates, possibly time-dependent, $b_{li}$ still represents the vector of patient-specific parameters (but now patient $i$ is nested within the level 3 cluster $l$), and $c_l$ represents the vector of level 3 parameters such that $c_l \sim N(0, \Sigma_c)$. The corresponding specification of the event submodel may take the form

$$h_{li}(t) = h_0(t) exp\,(v'_{li}(t)\gamma + \alpha\,\mu_{li}(t)) \qquad (9)$$

Because the additional clustering occurs at a level in the hierarchy that is higher than the patient we can simply use an association structure based on the patient-level expected value of the longitudinal outcome, without any need to derive a summary quantity based on lower-level units. The specification in (9) would assume that the hazard of the event for patient $i$ at time $t$ is associated with the patient-specific expected value of the longitudinal marker at time $t$, incorporating the effects of any higher level clustering. Note that the specification in (9) could be easily extended to any other patient-level function of the longitudinal submodel parameters, such as the patient-specific rate of change in the marker (i.e. slope) at time $t$ or the area under the patient-specific marker trajectory (i.e. integral) up to time $t$.

A possible extension would be to include a shared frailty term in the event submodel

$$h_{li}(t) = h_0(t) exp\ (v'_{li}(t)\gamma + \alpha\ \mu_{li}(t) + \delta_l) \qquad (10)$$

where $\delta_l$ is, for example, assumed to follow a normal or log-Gamma distribution. The inclusion of the shared frailty term does not induce an association with the longitudinal submodel, but it does allow for correlation in the event times of patients within a level 3 cluster. Note that if the variance of the $\delta_l$ parameters is close to zero, then this would suggest there is little within-cluster correlation in the event times and the shared frailty term could be dropped from the model. Moreover, if the number of level 3 groups was small then another alternative would be to include the level 3 group as a fixed effect covariate in the event submodel or as a stratification factor for the baseline hazard. The benefit of these latter models is that they may be computationally simpler than specifying a shared frailty term as a random effect.

6.2 Model formulation based on a *higher*-level association structure

An alternative possibility is that the hazard of the event for patient $i$ need only be related to the higher-level cluster's deviation from the average. That is, we can consider a shared random effects joint model of the form

$$h_{li}(t) = h_0(t) exp\ (v'_{li}(t)\gamma + \alpha\ c_l) \qquad (11)$$

where $c_l$ might, for example, represent the clinic-level random intercept. In this case, we would have a model in which we assume that the hazard of the event for patient $i$ is associated with the way in which their clinic's biomarker measurements deviate from the average clinic, but not with any time-varying characteristics of the patient themselves. Here, the random effect $c_l$ serves two purposes in the event submodel. First, it allows for within-cluster correlation in the event times (as previously described for the shared frailty). Second, it allows for dependence between the event and longitudinal processes through a shared parameter at the level of clustering factor $l$.

7. Discussion

Increasingly complex data structures are being accommodated under a joint longitudinal and time-to-event modelling framework. In this paper we have described a new joint modelling approach that allows for multilevel hierarchical data, where the data structure includes clustering factors beyond that of the individual. Such data structures commonly appear in clinical and epidemiological research, however, they have not previously been incorporated into a joint modelling framework. Standard joint modelling approaches aim to model patient-level measurements of a clinical biomarker, however, greater flexibility can be achieved by incorporating both patient-specific and cluster-specific effects in the longitudinal submodel when those levels of clustering are present in the underlying data structure. Moreover, it allows an additional set of association structures to be used for modelling the association between the longitudinal biomarker and the patient-level risk of the event. We proposed a set of possible association structures that could be used in most settings, however, the most appropriate choice of association structure is likely to depend on the primary research question and data structure that is relevant to the application at hand. By incorporating the multilevel structure into our joint modelling approach, we are able to formulate a model that answers the research question appropriately. For instance, in our application, patient-level summaries of the lesion-specific trajectories are likely to be meaningful in a way that quantities obtained by ignoring the lesion level would not be.

A potential limitation of the modelling in our application is that the observed event times were subject to interval censoring. This interval censoring is evident from the "steps" that can be seen in the Kaplan-Meier curve in Figure 2. This is due to clinicians in the IPASS trial declaring disease progression at the scheduled clinic visits. In our application we ignored this interval censoring and so an avenue for future work will be to accommodate this interval censoring within our proposed joint modelling framework. Moreover, in future work, we would like to separately assess the competing event outcomes of death and disease progression by considering cause-specific competing risks event submodels. In this way, we will be able to separate out the cause-specific associations between tumor burden and each of the competing events.

A significant strength of this paper is that our proposed model, described in Section 3, has been implemented as part of the rstanarm R package. A benefit of having implemented this model as part of that package is that researchers can easily fit the model to their data, via a user-friendly interface with customary R formula syntax and data frames. The back-end estimation of the model is carried out under a full Bayesian specification with priors on all unknown parameters. A variety of prior distributions are available to the user, as well as a variety of exponential family and link function options for the longitudinal outcome, thereby providing significant flexibility. In addition, the package allows users to estimate a joint model with multiple longitudinal outcomes (i.e. a multivariate joint model) of which one or more can have the multilevel structure described in Section 3. We hope that by providing user-friendly software and example code (Supplementary Materials) for fitting the proposed model, we will help to facilitate its use in a wide variety of applications.

Figure 1. Example of the hierarchical structure of joint model data under three possible scenarios: (a) one longitudinal biomarker (tumor size) where the patient is the only clustering factor; (b) one longitudinal biomarker (tumor size) where there are two clustering factors (lesions clustered within patients); (c) two longitudinal biomarkers (tumor size and circulating DNA), one of which has one clustering factor (the patient), and one of which has two clustering factors (lesions clustered within patients).

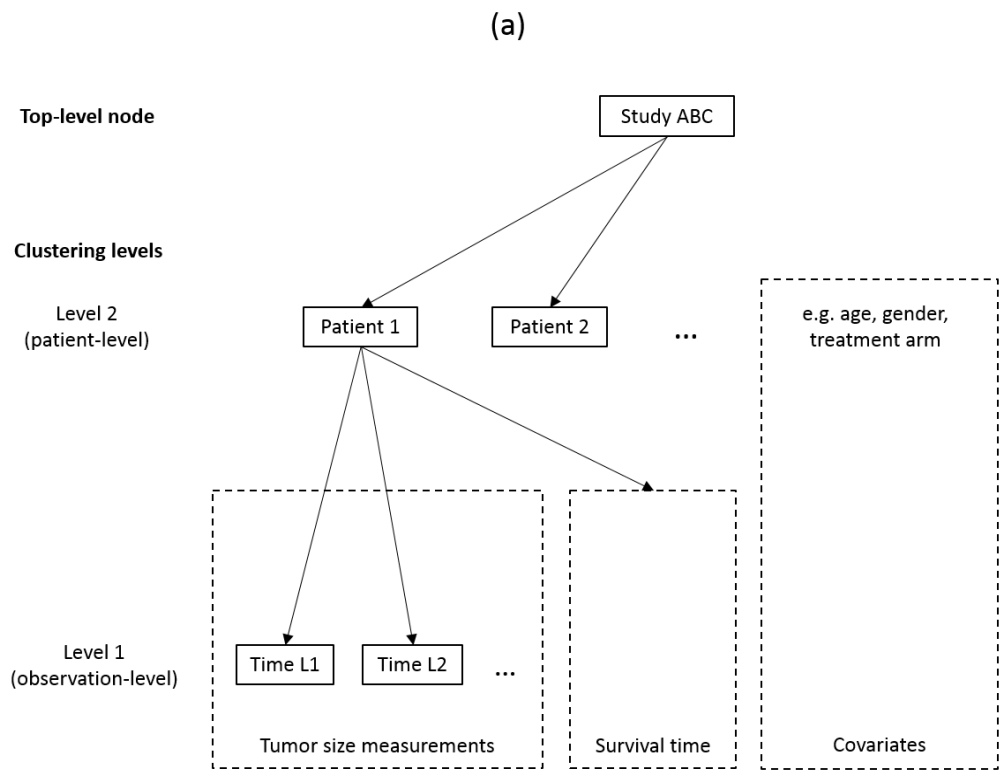

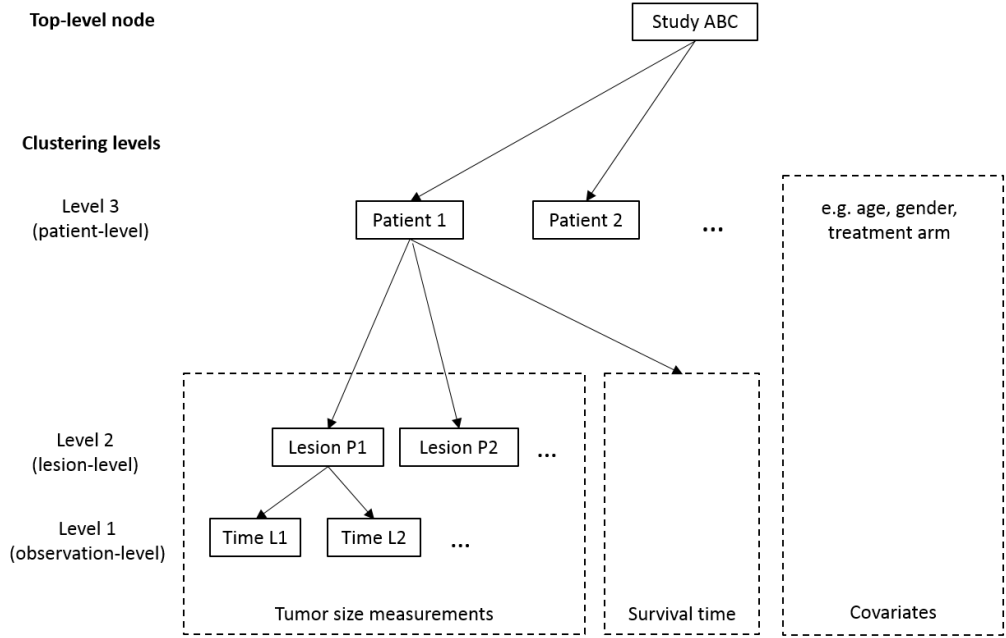

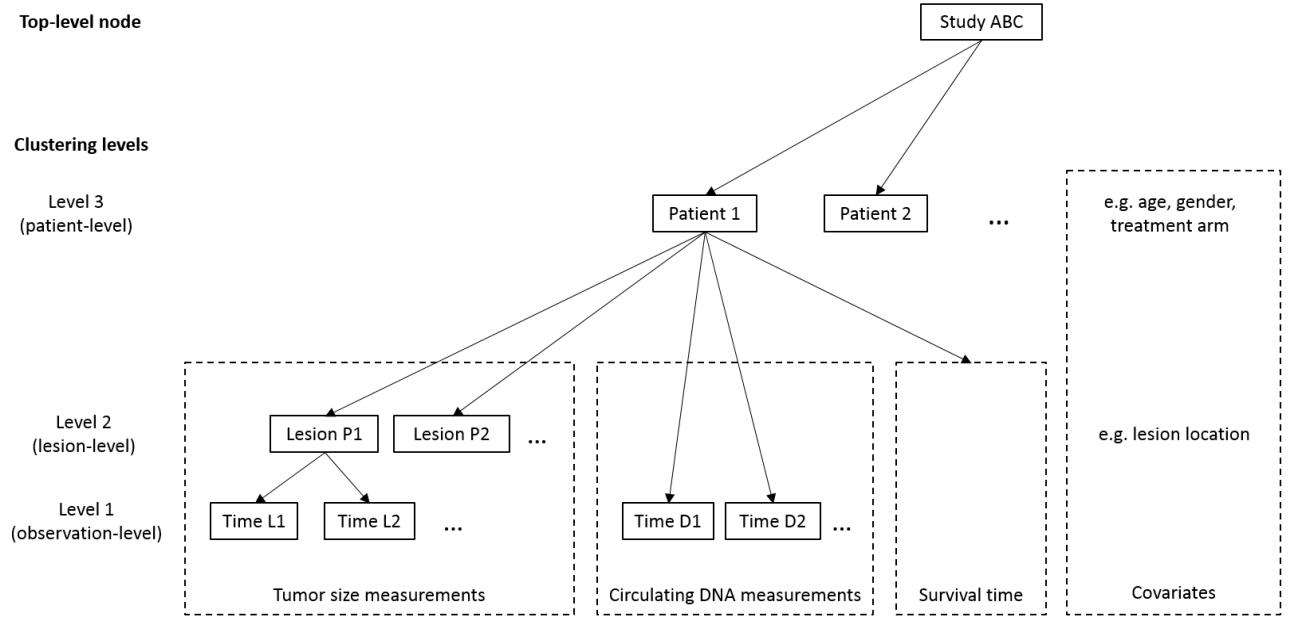

Figure 2. Overall Kaplan-Meier curve for progression-free survival. The values provided at the top of the plot are the numbers of patients still at risk for the event.

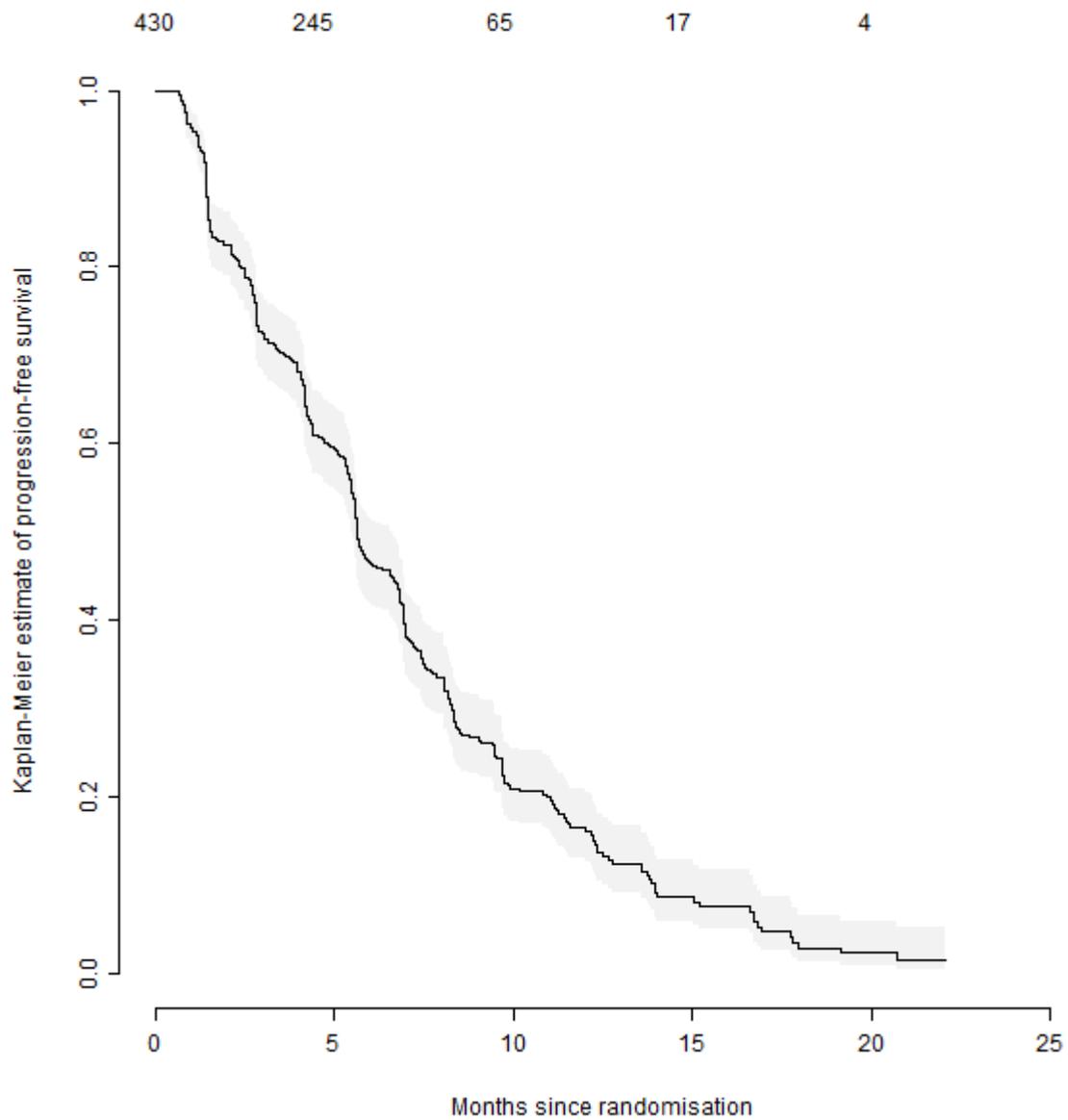

Figure 3. Observed longitudinal biomarker measurements (longest diameter of the lesion) and the fitted lesion-specific longitudinal trajectories (with 95% prediction intervals) under the joint model, for a selection of patients. Each panel of the figure shows a different patient, with some patients having multiple lesions. The dashed vertical line shows each patient's event or censoring time.

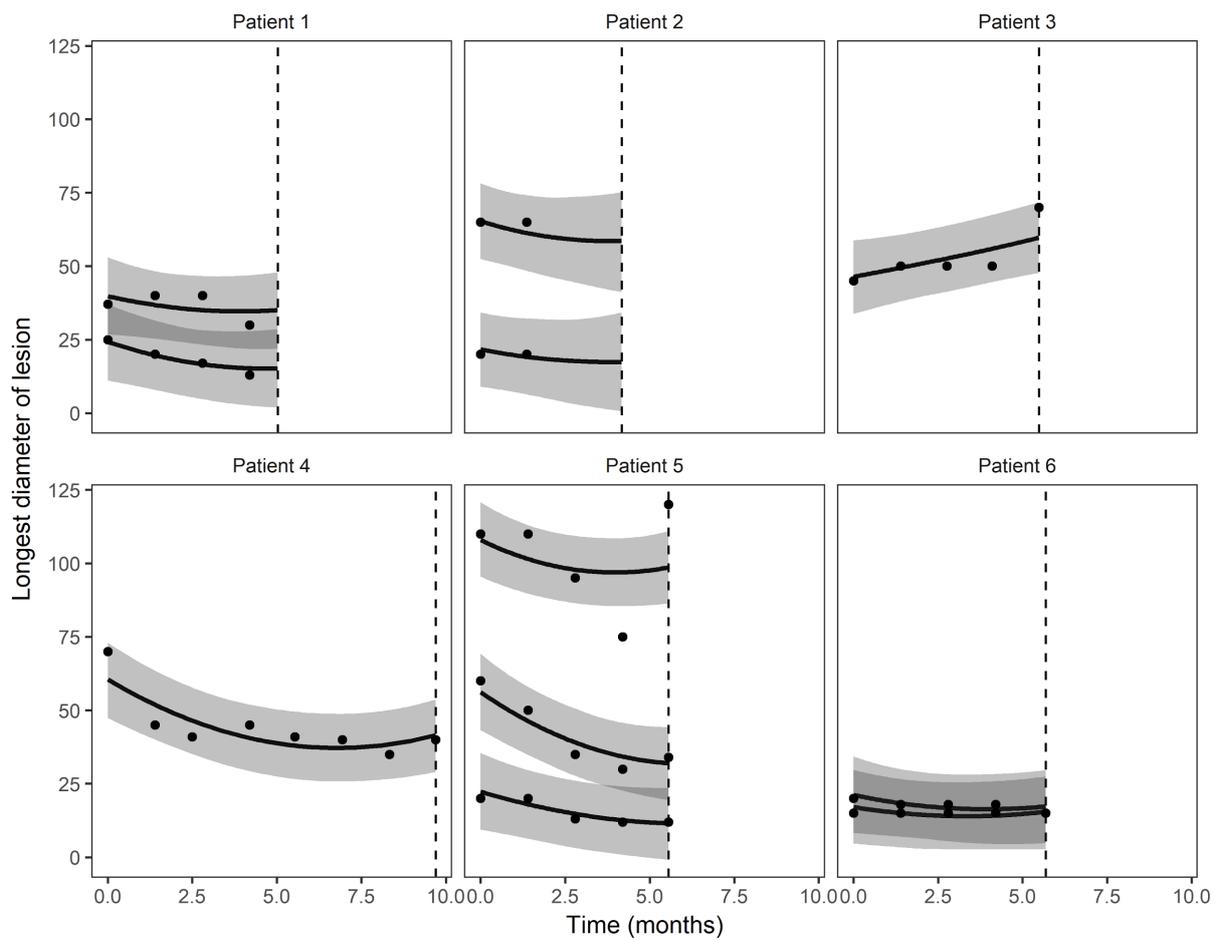

Table 1. Estimated time-dependent AUC for the proposed joint model using various association structures. The AUC is calculated using a landmark time of $t = 5$ months and horizon time of $t = 10$ months.

| Association structure | Time-dependent AUC |
|---|---|
| No biomarker data (i.e. no association structure) | 0.50 |
| Lesion-specific value | |
|     Sum | 0.62 |
|     Average | 0.56 |
|     Maximum | 0.61 |
|     Minimum | 0.55 |
| Lesion-specific value & slope | |
|     Sum | 0.65 |
|     Average | 0.64 |
|     Maximum | 0.66 |
|     Minimum | 0.59 |

Abbreviations. AUC: area under the (receiver operating characteristic) curve.

Table 2. Fixed effect parameter estimates (posterior means and 95% credible interval limits) from the joint model. The estimates for the event submodel are hazard ratios and the coefficients for the B-splines baseline hazard have been omitted.

| Parameter | Estimate | Lower | Upper |
|---|---:|---:|---:|
| Longitudinal submodel | | | |
|   Intercept | 23.0 | 21.3 | 24.7 |
|   Group (ref: EGFR+) | | | |
|     EGFR-, carboplatin plus paclitaxel | 4.0 | 0.8 | 7.1 |
|     EGFR-, gefitinib | 16.9 | 13.2 | 20.4 |
|   Time effects | | | |
|     Linear term (orthogonalised) | -0.1 | -73.3 | 76.7 |
|     Quadratic term (orthogonalised) | 450.3 | 391.6 | 512.5 |
|   Group * Linear interaction | | | |
|     EGFR-, carboplatin plus paclitaxel * Linear | 315.2 | 195.1 | 438.4 |
|     EGFR-, gefitinib * Linear | 390.0 | 127.5 | 660.4 |
|   Group * Quadratic interaction | | | |
|     EGFR-, carboplatin plus paclitaxel * Quadratic | 23.7 | -74.3 | 123.4 |
|     EGFR-, gefitinib * Quadratic | -524.8 | -697.0 | -351.1 |
| Event submodel | | | |
|   Physical functioning (ref: in bed >50% of the time) | | | |
|     Normal activity | 0.6 | 0.4 | 1.0 |
|     Restricted activity | 0.6 | 0.4 | 1.0 |
|   Association parameters (exponentiated) | | | |
|     Value (diameter of largest lesion at time $t$) | 1.011 | 1.004 | 1.017 |
|     Slope (rate of change in fastest growing lesion at time $t$) | 1.56 | 1.42 | 1.75 |

Abbreviations. ref: reference category; EGFR: epidermal growth factor receptor (mutation status).

**Supplementary materials for "Joint longitudinal and time-to-event models for multilevel hierarchical data"**

Samuel L Brilleman[1,2] *

Michael J Crowther[3]

Margarita Moreno-Betancur[2,4,5]

Jacqueline Buros Novik[6]

James Dunyak[7]

Nidal Al-Huniti[7]

Robert Fox[7]

Jeff Hammerbacher[6,8]

Rory Wolfe[1,2]

**Author Affiliations:** [1] Department of Epidemiology and Preventive Medicine, School of Public Health and Preventive Medicine, Monash University, Melbourne, Australia; [2] Victorian Centre for Biostatistics (ViCBiostat), Melbourne, Australia; [3] Biostatistics Research Group, Department of Health Sciences, University of Leicester, Leicester, UK; [4] Clinical Epidemiology and Biostatistics Unit, Murdoch Children's Research Institute, Melbourne, Australia; [5] Melbourne School of Population and Global Health, University of Melbourne, Melbourne, Australia; [6] Department of Genetics and Genomic Sciences, Icahn School of Medicine at Mount Sinai, New York, NY, USA; [7] Quantitative Clinical Pharmacology, AstraZeneca, Waltham, MA, USA; [8] Department of Microbiology and Immunology, Medical University of South Carolina, Charleston, SC, USA;

* Corresponding author:

Postal: School of Public Health and Preventive Medicine, Monash University, 553 St Kilda Road, Melbourne, VIC 3004, Australia

Phone: +61 (4) 9903 0802

Email: sam.brilleman@monash.edu

1. Further details on the model estimation

Our Bayesian specification requires prior distributions on all unknown parameters. We refer the reader to the documentation of the rstanarm R package for details on prior distributions, since our model was estimated using default priors implemented in the package. In brief, we used weakly informative normal distributions for each of the regression coefficients (fixed effects). The residual standard deviation (for the longitudinal outcome) was given a weakly informative half-Cauchy distribution. The B-spline coefficients for the log baseline hazard were given weakly informative Cauchy distributions. The weakly informative priors were only intended to reduce support given to values of the parameters that would seem implausible based on the scale of magnitude of the data. They were *not* intended to provide support to specific parameter values based on prior knowledge or expert opinion.

For estimation of the model parameters we ran four MCMC chains in parallel, each with 1000 sample iterations preceded by a warm up period of 1000 iterations (i.e. 2000 iterations in total, of which 50% were warm up). Although this number of iterations would seem small for a complex model estimated using a Gibbs sampler, the estimation in Stan is based on a Hamiltonian Monte Carlo (HMC) algorithm, not Gibbs sampling. The HMC results in much lower autocorrelation between subsequent MCMC draws compared with Gibbs sampling, and therefore is much more efficient in terms of the effective sample size per iteration. For example, for each of the models with an association structure based on the expected value, the effective sample size for the estimated association parameter was 4000.

A potential limitation of the proposed approach is the additional computational complexity. Additional clustering factors mean that there are an increasing number of cluster-specific parameters (i.e. random effects) to be estimated and therefore computation time increases. In our application with 430 patients having a total of 1209 lesions, there were 430 patient-specific parameters (intercept only) and 3627 lesion-specific parameters (intercept and two polynomial terms) to be estimated. Computation time for the models with an association structure based on the expected value ranged between 1.5 and 3 hours. The differences in computation time were partly related to the random nature of the different MCMC chains, and partly related to the type of summary function used in the association structure (i.e. the sum, average, maximum, or minimum of the level 2 clusters). The type of association structure is of course part of the model definition and therefore the choice of association structure will have an influence on the shape of the target posterior distribution, with some resulting posteriors easier for the MCMC sampler to explore (i.e. less extreme curvature in the posterior). When the association structure was based on both the expected value and the slope, the computation times were slightly longer; ranging between 2 and 5.5 hours. These times are based on 1000 warm up iterations, followed by 1000 sample iterations, on a standard quad-core desktop with a 3.30GHz processor and 8GB RAM.

2. Example code for fitting the model

The model in the paper can be easily estimated after downloading the rstanarm R package from the Comprehensive R Archive Network (CRAN). To download and install rstanarm, type the following into your R console:

```
> install.packages("rstanarm")
```

And then an example of the code used to fit the model presented in Table 2 of the main manuscript would be:

```
> library(rstanarm)

> mod <- stan_jm(
  formulaLong = ldiam ~
    cat * poly(months, degree = 2) +
    (poly(months, degree = 2) | lesid_usubjid) +
    (1 | usubjid),
  dataLong = ipass$lesions,
  formulaEvent = Surv(progmnth, censor_p) ~ whostat,
  dataEvent = ipass$surv,
  seed = 9837355, time_var = "months", id_var = "usubjid",
  assoc = c("etavalue", "etaslope"), grp_assoc = "max")
```

Where `ipass$lesions` is a data frame containing the outcome and covariate data for the longitudinal submodel, and `ipass$surv` is a data frame containing the outcome and covariate data for the event submodel. Unfortunately, since the IPASS data used in the application in the main manuscript is not publically available, we cannot provide the reader with these data frames.